\documentclass[aps, prl,twocolumn,superscriptaddress,groupedaddress]{revtex4}

\usepackage{graphicx}% Include figure files

\begin{document}

\title {Colossal non-saturating linear magnetoresistance in two-dimensional electron systems at a GaAs/AlGaAs heterointerface}

\author{M. A. Aamir}
\email{aamir@physics.iisc.ernet.in}
\affiliation{Department of Physics, Indian Institute of Science, Bangalore 560 012, India.}
\author{Srijit Goswami}
\affiliation{Department of Physics, Indian Institute of Science, Bangalore 560 012, India.}
\author{Matthias Baenninger}
\affiliation{Cavendish Laboratory, University of Cambridge, J.J. Thomson Avenue, Cambridge CB3 0HE, United Kingdom.}
\author{Vikram Tripathi}
\affiliation{Department of Theoretical Physics, Tata Institute of Fundamental Research, Homi Bhabha Road, Mumbai 400005, India}
\author{Michael~Pepper}
\affiliation{Department of Electrical and Electronic Engineering, University College, London WC1E 7JE, United Kingdom}
\author{Ian~Farrer}
\affiliation{Cavendish Laboratory, University of Cambridge, J.J. Thomson Avenue, Cambridge CB3 0HE, United Kingdom.}
\author{David~A.~Ritchie}
\affiliation{Cavendish Laboratory, University of Cambridge, J.J. Thomson Avenue, Cambridge CB3 0HE, United Kingdom.}
\author{Arindam~Ghosh}
\affiliation{Department of Physics, Indian Institute of Science, Bangalore 560 012, India.}

%\date{\today}

\begin{abstract}

Engineering devices with a large electrical response to magnetic field is of fundamental importance for a range of applications such as magnetic field sensing and magnetic read-heads. We show that a colossal non-saturating linear magnetoresistance (NLMR) arises in two-dimensional electron systems hosted in a GaAs/AlGaAs heterostructure in the strongly insulating regime. When operated at high source-drain bias, the magnetoresistance of our devices increases almost linearly with magnetic field reaching nearly 10,000\% at 8~Tesla, thus surpassing many known non-magnetic materials that exhibit giant NLMR. The temperature dependence and mobility analysis indicate that the NLMR has a purely classical origin, driven by nanoscale inhomogeneities. A large NLMR combined with small device dimensions makes these systems a new and attractive candidate for on-chip magnetic field sensing.

\end{abstract}

%/PACS:  75.47.Gk,  73.40.-c,  75.70.-i

\maketitle

%\It may seem that a magnetic nature of the material is vital for this~\cite{Berkowitz, Xiao, Baibich}, but some materials exhibit it even without magnetism at any scale. Over the last decade, it has been shown that disorder alone can provide a route to large magnetoresistance (MR) in non-magnetic semiconductors~\cite{Xu,Meera,MeeraPRB, Delmo, Johnson}.

In a non-magnetic semiconductor, giant positive linear non-saturating magnetoresistance (NLMR) can have both classical~\cite{Xu,Delmo,Jingshi,Meera} and quantum origin~\cite{Abrikosov, Abrikosov_layered}. The quantum NLMR, originally proposed by Abrikosov~\cite{Abrikosov, Abrikosov_layered}, is applicable in the extreme quantum limit where $\hbar\omega_c \gg E_F, k_BT$ ($\omega_c$ and $E_F$ are the cyclotron frequency and Fermi energy, respectively). This criterion can be attained in a restricted class of materials which include semimetals such as bismuth, or narrow band gap semiconductors with very low effective mass (\emph{e.g.} InSb~\cite{Jingshi}, graphene~\citep{Graphene}, and topological insultors~\cite{TI}). The classical NLMR, on the other hand, is commonly observed in systems with an inhomogeneous carrier (and hence mobility) distribution~\cite{Meera,MeeraPRB}. It is a purely geometric effect, where, in the presence of transverse magnetic field, a misalignment between current paths and the externally applied bias mixes the off-diagonal components of the magnetoresistivity tensor, resulting in a NLMR. Several classically inhomogeneous conductors, most notably the mildly doped silver chalcogenides (Ag$_{2+\delta}$Se or Ag$_{2+\delta}$Te)~\cite{Xu} and InSb polycrystals~\cite{Jingshi}, display extremely large NLMR, where the inhomogeneity is associated with intrinsic disorder such as grain boundaries, dopant clustering etc. Thus a handle on the disorder, both in magnitude and length scale, could yield a new class of high sensitivity magnetoresistive devices. However, achieving such a control in bulk materials is not a trivial task.

Two-dimensional electron systems (2DESs) in semiconductor multi-layers, in particular doped GaAs/AlGaAs heterostructures, offer a material platform in which disorder can be tuned with great precision using molecular beam epitaxy. At high carrier density, since the hetero-interface is physically separated from the ionized dopants, it is a homogeneous medium which hosts high mobility electrons. Therefore, it is not expected to be a good candidate for exhibiting NLMR. This is confirmed by numerous magnetoresistance (MR) measurements in 2DES, which are best studied in two limits.  At high carriers densities, the MR of a 2DES is oscillatory in magnetic field ($B$) due to the Shubhnikov-de Haas effect, while at lower carrier densities, at the onset of localization, the MR increases exponentially with $B$ due to variable range hopping~\cite{Matthias}. Clearly, studies in neither regime have thus far revealed a NLMR.

\begin{figure*}

\includegraphics[scale=0.24]{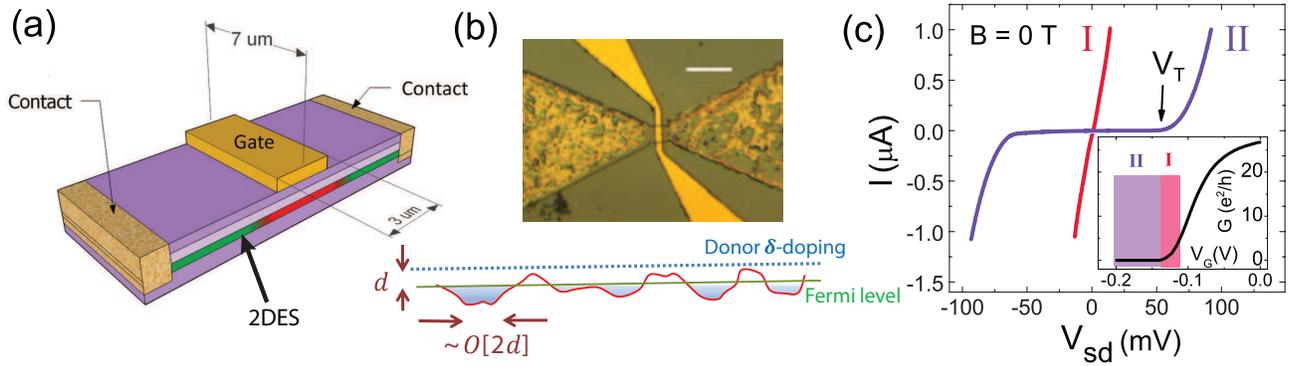}
\vspace{-15pt}
\caption{(a) Schematic of the device --- the gate modulates the electron number density only in region of 2DES below it (highlighted red) thus defining the mesoscopic region of interest (b) Top: Optical micrograph of the device. Scale bar is 20~$\mu$m. Bottom: Potential landscape profile for the electrons at very low number density (c) Inset: Equilibrium conductance as a function of the gate voltage.  Region I represents gate voltage just above the pinch-off (the point when equilibrium conductance becomes zero); Region II represents another set of gate voltages below the pinch-off point. The main figure shows $I-V$ characteristics of the device for these two distinct regions at 0.3~K. The arrow indicates the threshold voltage $V_T$.
}
\end{figure*}

A 2DES however does become inhomogenous when it is depleted with a strong negative gate voltage ($V_g$)~\cite{Houten92}. This constitutes the backbone of our experiments in the following way: for a typical doped GaAs/AlGaAs heterostructure with spacer thickness $d$, the Coulomb potential from the randomly scattered ionized dopants form the dominant component of disorder. This causes the conduction band minimum to fluctuate as a function of position (see Fig.~1b) with a correlation length of $\sim 2d$~\cite{Tripathi06,Tripathi07,Neilson}.  In this regime, the 2DES disintegrates into small puddles of charge that often manifest in Coulomb blockade effects in mesoscopic devices~\cite{JPhys_Arindam, Houten92,Tripathi07}. In essence, this increase in inhomogeneity arises due to the weakening of electrostatic screening of the background disorder potential landscape, causing the spatial fluctuations in carrier density to be of the same order as the carrier density itself. We show that such a strong density variation at the GaAs/AlGaAs interface does in fact give rise to a giant NLMR in a simple gate-tunable mesoscopic system.

%Thus when $E_F$ is lowered with a negative $V_g$ to the order of conduction band fluctuations, the charge distribution becomes inhomogeneous at the the scale $\sim 80 - 100$~nm for a $d = 40$~nm heterostructure, similar to that of silver chalcogenides.\\

\begin{figure}[b]

\includegraphics[width=0.78\linewidth]{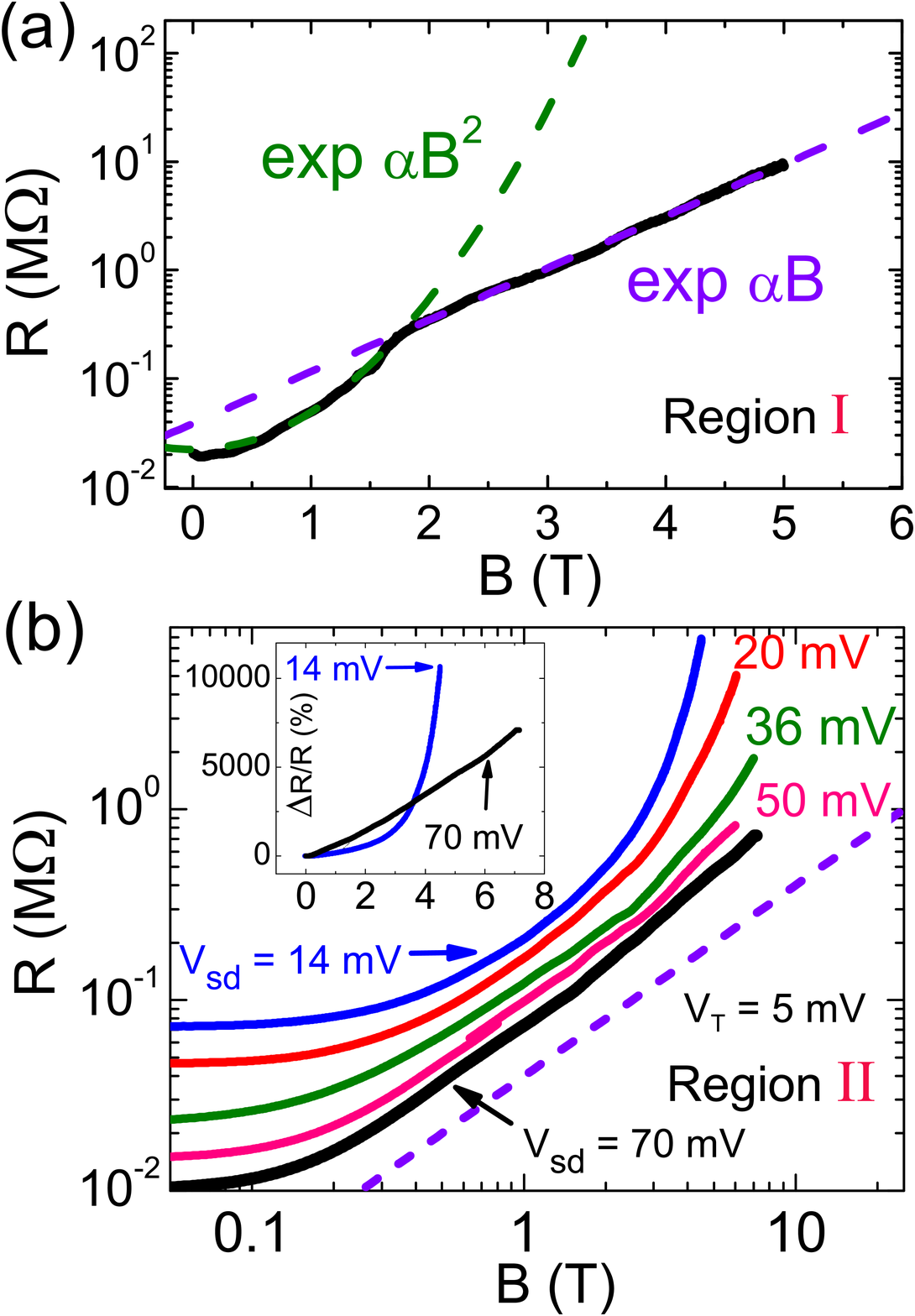}
\caption{ (a) MR of the sample for gate voltages above the pinch-off voltage (Region I) obtained using ac lock-in measurements (b) MR at different $V_{sd}$ below the pinch-off point (Region II). As source-drain is increased, the MR gradually transforms from a rapidly rising curve a to power law with exponent 1.1. The dashed line is $R \propto B$. Inset: Percentage change in resistance $\Delta R/R~(\%)$ for two $V_{sd}$ of 14~mV and 70~mV}

\end{figure}

\begin{figure*}

\includegraphics[scale=0.26]{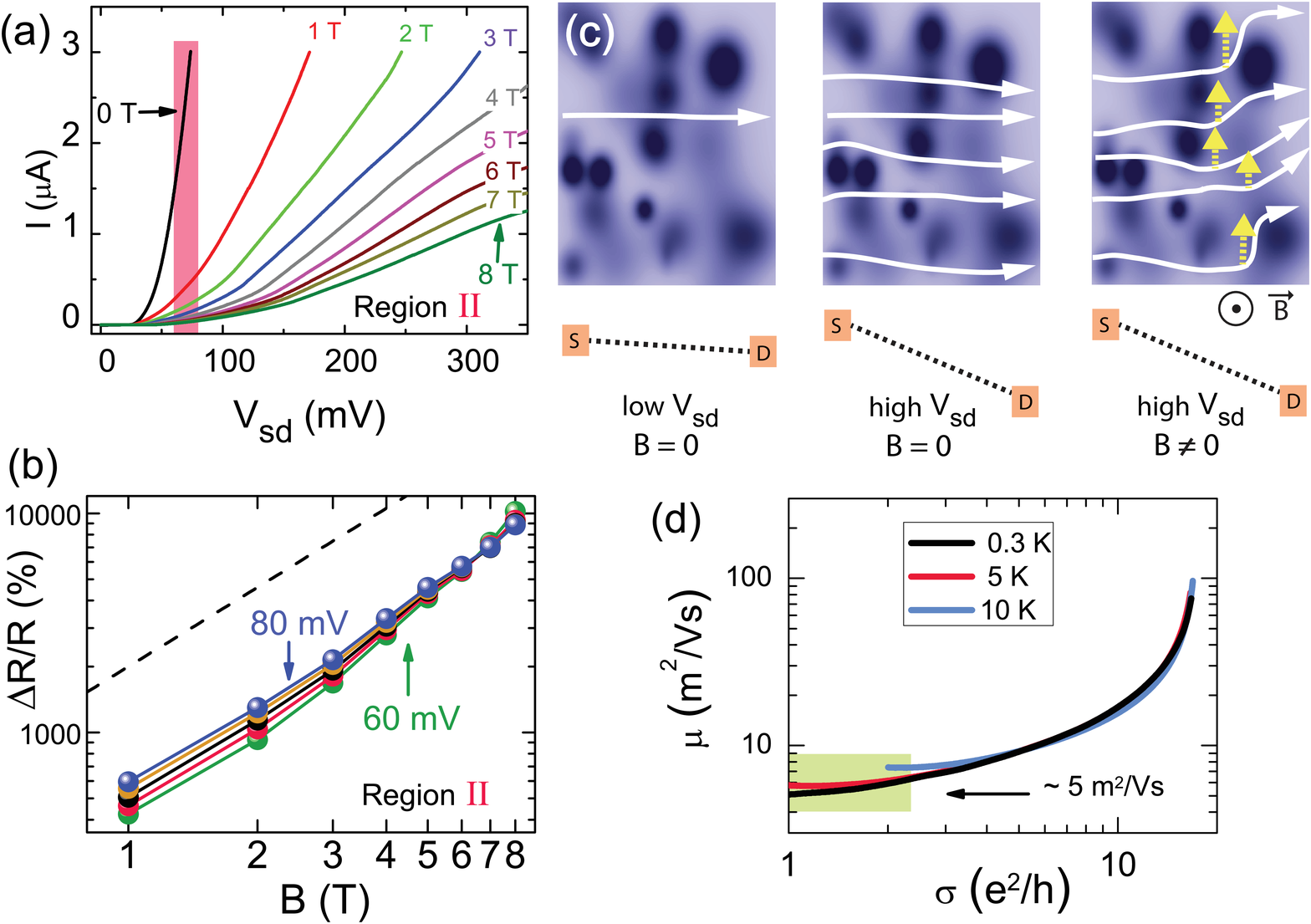} %earlier scale 0.25
\vspace{0pt}
\caption{(a) $I-V_{sd}$ characteristics as a function of magnetic field from 0 T to 8 T (in steps of 1 T) at 0.3 K (Region II) (b) $\Delta R/R~(\%)$ for different $V_{sd}$ chosen from the shaded region of (a) ($V_{sd}$ step size is 5 mV) (c) A schematic illustrating the possible origin of NLMR: The light (dark) regions represent areas of high (low) carrier number density. White arrows represent the current paths. Left: At low $V_{sd}$, there is only one or very few current channels. Middle: For high $V_{sd}$, the number of current paths proliferate rapidly. Right: On applying a perpendicular magnetic field at high $V_{sd}$, the plurality of the paths in the disordered medium gives rise to non-trivial current trajectories which have significant transverse component (highlighted by dotted yellow arrows) that mixes the Hall contribution into the longitudinal MR, thus giving rise to NLMR (d) The mobility (calculated using Drude's formula) as a function of the conductivity of the device at 0.3~K, 5~K and 10~K.}

\end{figure*}

Our experiments are carried out in strongly insulating 2DES's in the GaAs/AlGaAs heterostructure where we explore the magnetic response of the system as a function of gate voltage ($V_g$) and drain-source bias ($V_{sd}$). The heterostructures contain a $\delta$-doped layer of Si dopants of doping density $2.5\times10^{12}$~cm$^{-2}$ placed 40~nm above the GaAs/AlGaAs interface. The 2DES lies 90 nm below the surface, with an as-grown mobility $\sim  300$~m$^2$/Vs. The etched mesa and a surface gate define the effective geometry of the device. Here we present results from a $3~\mu$m$ \times 7~\mu$m device (see Fig.~1a and Fig.~1b), though similar devices on two separate wafers showed qualitatively similar results (see Supplementary Material). The gate is used to pinch the device off by application of a negative voltage, and the resulting conductance curve is shown in the inset of Fig.~1c. Fig.~1c shows $I-V_{sd}$ characteristics at two values of $V_g$, representative of the two regions identified as Regions I and II. Region I represents the onset of localization at linear conductance $G \sim e^2/h$, and has a reasonably ohmic $I-V_{sd}$. Region II however lies deep in the pinched-off regime where the 2DES is expected to become inhomogeneous. Here, we find the $I-V_{sd}$ characteristics are strongly non-linear with an insulating regime up to a threshold voltage $V_{T}$ (indicated by an arrow in Fig.~1c), which increases monotonically as $V_g$ is made increasingly negative (see Supplementary Material). Within this threshold, the current is immeasurably small ($< 10^{-11}$~A), and conduction sets in rapidly only after the source-drain bias is increased beyond $V_T$. This rapid rise of current above threshold cannot be associated with avalanche breakdown common in semiconductors~\cite{Schoonus, Sun, Ciccarelli}. This is because the applied electric fields are 3 orders of magnitude lower than the breakdown field in GaAs~\cite{Sun}. The strong dependence of $I-V_{sd}$ curves on magnetic field (to be discussed later) also eliminates self-heating effects as a cause for the rapidly rising current. In fact, a systematic analysis of the $I-V_{sd}$ characteristics reveals that they follow a power law, characteristic of a disordered array of charge puddles (see Supplementary Material). This provides direct evidence that in Region II the system is indeed highly inhomogeneous.

This inhomogeneity has a dramatic effect on the MR when subjected to a transverse magnetic field. At the onset of localization (prior to pinch-off), represented by Region I in Fig.~1c, the resistance R (obtained using equilibrium measurements) increases exponentially with $B$. As shown in Fig.~2a,  $R \sim \exp(B^2)$ at low $B$, which changes to $R \sim \exp(B)$ as $B$ exceeds 2~T. These are characteristic features of hopping transport in the perturbative (low-$B$) and non-perturbative (high-$B$) regime, and recently studied in detail by some of us~\cite{Matthias}. In the sub pinch-off regime, represented by Region II, $ R $ was found to behave very differently. Most notably, its structure is highly sensitive to the magnitude of $V_{sd}$. To elucidate this, we define the dc magnetoresistance $R = R(V_{sd},B) = V_{sd}/I(V_{sd},B)$, and plot it as a function of $B$ for different values of $V_{sd}$ in Fig.~2b (we carried out the same analysis with differential resistance $dV_{sd}/dI(V_{sd},B)$, which did not yield significantly different result). Strikingly, the variation in $R$ weakens with increasing $V_{sd}$, and for sufficiently high $V_{sd} >> V_T$ ($V_T \approx 5 $~mV at $V_g=-0.174$~V, $B = 0$~T), $R$ increases {\it linearly} with $B$. In order to quantify the change in R, we evaluate the percentage change 

\[\frac{\Delta R}{R} (\%) = \frac{[R(V_{sd},B) - R(V_{sd},0)]}{R(V_{sd},0)} \times 100\%\]

 This quantity has been plotted for the two extremal $V_{sd}$ in the inset of Fig.~2b, which clearly shows that even the percentage change in $R$ has transformed from a rapidly rising exponential curve to a strikingly linear form.

To study this more systematically, we look at the evolution of the $I-V_{sd}$ characteristics as $B$ is increased from 0 to 8~T. We have chosen a gate voltage corresponding to $V_T\approx 30$~mV (see Supplementary Material). Fig.~3a shows that a non-zero $B$ suppresses the current drastically, leading to a positive MR that increases monotonically with $B$. Within the experimentally achievable $B$ (8~T) the measured $R$ did not show any sign of saturation. We have evaluated $\Delta R/R~(\%)$ (as defined above) for a few values of $V_{sd}$ above $V_T$ (the range is highlighted in Fig.~3a). As shown in Fig.~3b, the MR in our mesoscopic 2DES reaches almost 10,000\% at $B=8$~T for $V_{sd} = 80$~mV. As expected from Fig. 2b, the MR becomes nearly linear with increasing $V_{sd}$. The dashed line in Fig. 3b represents $\Delta R/R \propto B^{1.2}$. 

%The rise in MR, extracted from its value at 1~T, is an increasing function of $V_{sd}$ (see Supplementary Material), the highest of which that is directly observed is about 500\%  at $V=70$~mV\footnote{MR is extracted for $V_{sd}$ higher than $V=70$~mV by modeling the $I-V$ curve for $B=0$~T with a power law (see Supplementary Material)}.

The inherent inhomogeneity in charge distributions and a parabolic dispersion relation for the carriers makes the classical model for NLMR proposed by Parish and Littlewood~\cite{Meera,MeeraPRB} particularly applicable in our devices. The findings in Ref. 4 followed from a numerical simulation of an equivalent node-link network model. Using the same physical principles, we have presented in the Supplementary Material an alternate theoretical analysis for NLMR in an inhomogeneous conductor, which augments existing theoretical descriptions~\cite{Meera, MeeraPRB, Herring, Bergman, Stroud, Guttal}. The classical model requires current flow from source to drain to occur via multiple channels in order to realize the non-trivial magnetic response of NLMR. The process by which NLMR arises in our devices is depicted schematically in Fig.~3c. At low $V_{sd}$ (left schematic), there are very few electron channels for conduction which are not sufficient in number to give rise to NLMR. However, a high $V_{sd}$ (middle schematic) opens up many more conduction channels. This is similar to the non-equilibrium transport in disordered array of quantum dots where multiplication of paths is directly connected to the conduction threshold~\cite{Wingreen}. A perpendicular magnetic field distorts these current paths, which follows non-trivial trajectories through the inhomogeneous medium~\cite{Meera}, resulting in a substantial transverse component (dotted arrows in right schematic). This allows a significant mixing of the the off-diagonal components in the magneto-resistivity tensor, thus leading to the NLMR. This qualitative picture allows us to intuitively understand why we observe NLMR only at high values of $V_{sd}$. We note that studies on mildly doped Si also reported a seemingly similar dependence of the quantity $\Delta R/R$ on source-drain bias~\cite{Delmo}. However, the NLMR there was connected to an inhomogeneous electric field in the presence of space-charge injection. This scenario is certainly not applicable in our case since the bias applied in our experiments is significantly lower than that required to induce bulk semiconductor transport~\cite{Ciccarelli, Sun}.

\begin{figure}

\includegraphics[width=0.9\linewidth]{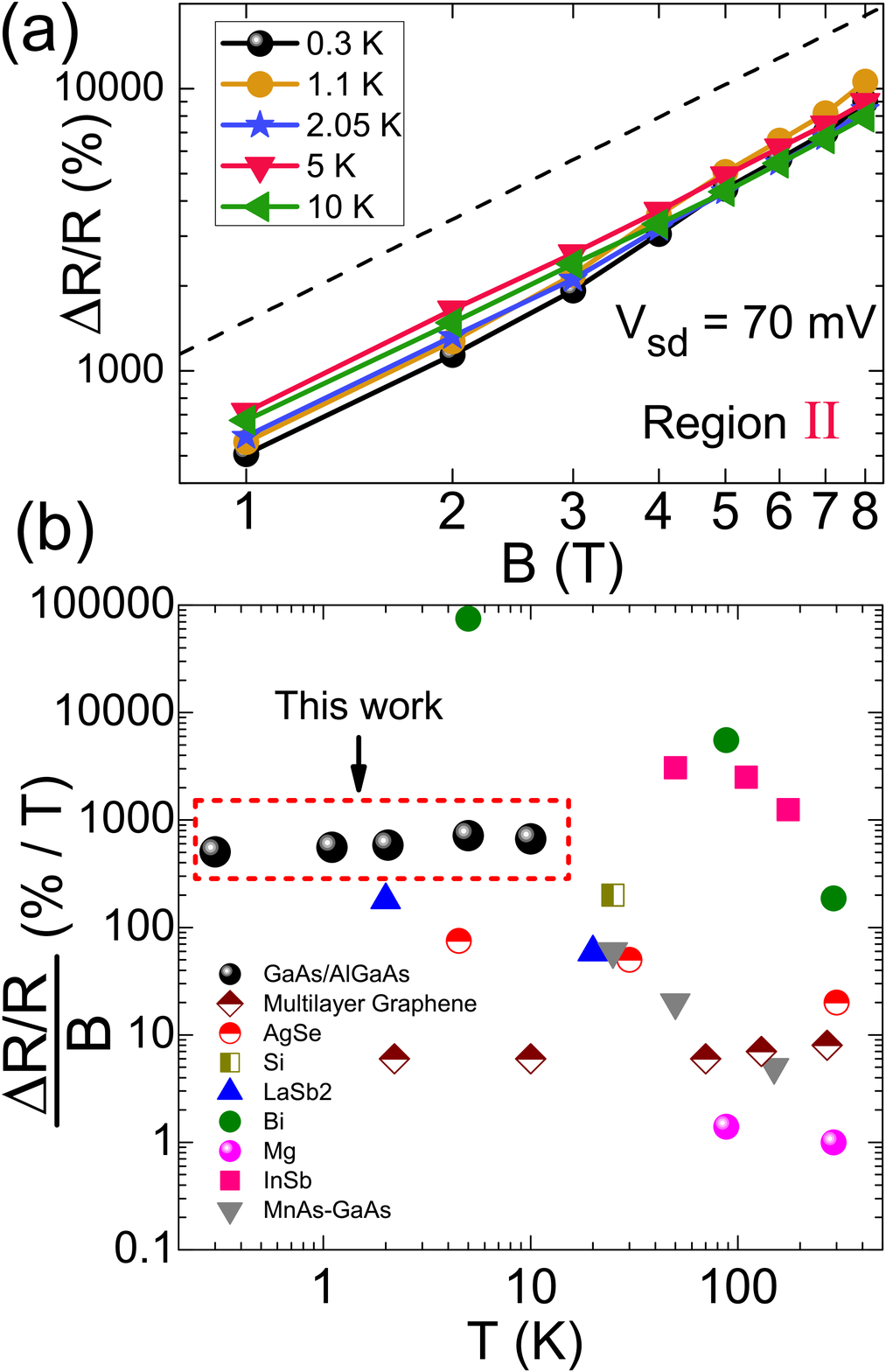}

\caption{(a) $\Delta R/R$ at different temperatures: 0.3~K, 1.1~K, 2.05~K, 5~K, 10~K for Region II. Dashed line represents $\Delta R/R \propto B^{1.2}$ (b) Comparison of MR of our device (highlighted by dashed box) with other non-magnetic materials which have linear MR: Ag$_{2+\delta}$Se~\cite{Xu}, Multilayer graphene~\cite{Graphene}, Bulk Si~\cite{Delmo}, InSb~\cite{Jingshi}, LaSb$_2$~\cite{Budko}, Bi~\cite{KapitzaBi, Yang}, Mg~\cite{Kapitza}, MnAs-GaAs composite~\cite{Johnson}.}

\end{figure}

An important question that must be addressed is whether the inhomogeneity in our systems can quantitatively explain the magnitude and field scales associated with the observed NLMR. Being a $\delta$-doped heterostructure, the dopant atoms are located at roughly the same distance from the 2DES, and hence the amplitude of conduction band fluctuations is unlikely to vary widely from one location to another. Therefore, it is not unreasonable to assume that the width of mobility distribution $\Delta\mu$ is smaller than the average mobility $\langle\mu\rangle$ among the charge puddles. In such a case, the classical model predicts a quadratic to linear MR transition at $B \sim \langle\mu\rangle^{-1}$, with $\Delta R(B)/R = c\langle\mu\rangle B$ in the linear regime~\cite{Guttal} ($c$ is a constant of order unity that depends on the details of mobility distribution). From Fig.~3(b), we find that $\Delta R(B)/R(0)B \approx 5$~T$^{-1}$, which is consistent with the fact that the transition to linear MR occurs around $0.1 - 0.2$~T (see Fig 2b). Moreover, a rough estimate of $\langle\mu\rangle$ can be obtained from the mobility of the device at the onset of localization which is usually identified by conductivity $\rightarrow e^2/h$ for a 2DES. Below this conductivity, the system starts becoming disintegrated and incompressible~\cite{Meir}. In Fig.~3d we have shown the variation in the Drude mobility of our device as a function of its conductivity for a few temperatures. The  limiting mobility at the onset of localization is about 5~m$^2$/Vs, which is in good order-of-magnitude agreement with $\Delta R(B)/R(0)B$. Interestingly, $\langle\mu\rangle$ is not a strong function of temperature in our heterostructures at least upto 10~K, presumably because electron-phonon interactions contribute less to the resistivity than elastic residual scattering. This should also carry over to a weak temperature dependence of the NLMR. Fig.~4a shows that, indeed, the NLMR retains roughly the same magnitude for a temperature range from 0.3~K to 10~K, thus spanning nearly two orders of magnitude.

%The appearence of NLMR only for $V > V_T$,  can be understood by combining the numerical results of Middleton \& Wingreen on current flow in a two dimensional array of quantum dots and Parish \& Littlewood model on magnetic response of random resistor network. The basic mechanism is represented with the cartoon in Fig.~3a. The classical model requires current flow from source to drain to occur in multiple channels in the resistor network in order to realize the nontrivial magnetic response in the form of NLMR. In a quantum dot array with charge disorder, the current flow occurs in a single optimal path when $V \lesssim V_T$. This correspons to the case of a single chain, 1 (row)$\times N$(columns), of random resistors which do not possess any nontrivial MR. The intrinsic MR of individual puddles or weak links are expected to dominate here.  Multiple current paths open up once $V > V_T$, which marks the onset of NLMR as well.

Finally, we compare the sensitivity of the disordered GaAs/AlGaAs heterostructure with other non-magnetic systems that exhibit NLMR. The sensitivity of our device is extracted from its value at 1~T in Fig. 4a and is an increasing function of $V_{sd}$ (see Supplementary Material). The highest value of which that is directly observed is about 500\% per Tesla  at $V_{sd}=70$~mV. In Fig.~4b we have compiled the reported values of $\Delta R(B)/R(0)B$ (in percentage per Tesla) from a wide variety of non-magnetic semiconductors and semimetals. Clearly, the sensitivity of the GaAs/AlGaAs system is comparable to existing non-magnetic materials. This, along with electrical tunability NLMR makes mesoscopic GaAs/AlGaAs heterostructures unique in comparison to other materials. In fact, more efficient and clever gating assemblies could possibly achieve an even higher sensitivity and form a new class of magnetoresistive sensors.

We would like to acknowledge support from Department of Science and Technology, Government of India and UKIERI. S.G. thanks the IISc Centenary Postdoctoral Fellowship. We are grateful to V. Venkataraman, Nigel Cooper and Peter Littlewood for useful discussions.


\begin{thebibliography}{29}
\expandafter\ifx\csname natexlab\endcsname\relax\def\natexlab#1{#1}\fi
\expandafter\ifx\csname bibnamefont\endcsname\relax
  \def\bibnamefont#1{#1}\fi
\expandafter\ifx\csname bibfnamefont\endcsname\relax
  \def\bibfnamefont#1{#1}\fi
\expandafter\ifx\csname citenamefont\endcsname\relax
  \def\citenamefont#1{#1}\fi
\expandafter\ifx\csname url\endcsname\relax
  \def\url#1{\texttt{#1}}\fi
\expandafter\ifx\csname urlprefix\endcsname\relax\def\urlprefix{URL }\fi
\providecommand{\bibinfo}[2]{#2}
\providecommand{\eprint}[2][]{\url{#2}}

\bibitem[{\citenamefont{Xu et~al.}(1997)\citenamefont{Xu, Husmann, Rosenbaum,
  Saboungi, Enderby, and Littlewood}}]{Xu}
\bibinfo{author}{\bibfnamefont{R.}~\bibnamefont{Xu}},
  \bibinfo{author}{\bibfnamefont{A.}~\bibnamefont{Husmann}},
  \bibinfo{author}{\bibfnamefont{T.~F.} \bibnamefont{Rosenbaum}},
  \bibinfo{author}{\bibfnamefont{M.}~\bibnamefont{Saboungi}},
  \bibinfo{author}{\bibfnamefont{J.~E.} \bibnamefont{Enderby}},
  \bibnamefont{and} \bibinfo{author}{\bibfnamefont{P.~B.}
  \bibnamefont{Littlewood}}, \bibinfo{journal}{Nature}
  \textbf{\bibinfo{volume}{390}}, \bibinfo{pages}{57} (\bibinfo{year}{1997}).

\bibitem[{\citenamefont{Delmo et~al.}(2009)\citenamefont{Delmo, Yamamoto,
  Kasai, Ono, and Kobayashi}}]{Delmo}
\bibinfo{author}{\bibfnamefont{M.~P.} \bibnamefont{Delmo}},
  \bibinfo{author}{\bibfnamefont{S.}~\bibnamefont{Yamamoto}},
  \bibinfo{author}{\bibfnamefont{S.}~\bibnamefont{Kasai}},
  \bibinfo{author}{\bibfnamefont{T.}~\bibnamefont{Ono}}, \bibnamefont{and}
  \bibinfo{author}{\bibfnamefont{K.}~\bibnamefont{Kobayashi}},
  \bibinfo{journal}{Nature} \textbf{\bibinfo{volume}{457}},
  \bibinfo{pages}{1112} (\bibinfo{year}{2009}).

\bibitem[{\citenamefont{Hu and Rosenbaum}(2008)}]{Jingshi}
\bibinfo{author}{\bibfnamefont{J.}~\bibnamefont{Hu}} \bibnamefont{and}
  \bibinfo{author}{\bibfnamefont{T.}~\bibnamefont{Rosenbaum}},
  \bibinfo{journal}{Nature Mater.} \textbf{\bibinfo{volume}{7}},
  \bibinfo{pages}{697} (\bibinfo{year}{2008}).

\bibitem[{\citenamefont{Parish and Littlewood}(2003)}]{Meera}
\bibinfo{author}{\bibfnamefont{M.~M.} \bibnamefont{Parish}} \bibnamefont{and}
  \bibinfo{author}{\bibfnamefont{P.~B.} \bibnamefont{Littlewood}},
  \bibinfo{journal}{Nature} \textbf{\bibinfo{volume}{58}}, \bibinfo{pages}{162}
  (\bibinfo{year}{2003}).

\bibitem[{\citenamefont{Abrikosov}(1998)}]{Abrikosov}
\bibinfo{author}{\bibfnamefont{A.~A.} \bibnamefont{Abrikosov}},
  \bibinfo{journal}{Phys. Rev. B} \textbf{\bibinfo{volume}{58}},
  \bibinfo{pages}{2788} (\bibinfo{year}{1998}).

\bibitem[{\citenamefont{Abrikosov}(1999)}]{Abrikosov_layered}
\bibinfo{author}{\bibfnamefont{A.~A.} \bibnamefont{Abrikosov}},
  \bibinfo{journal}{Phys. Rev. B} \textbf{\bibinfo{volume}{60}},
  \bibinfo{pages}{4231} (\bibinfo{year}{1999}).

\bibitem[{\citenamefont{Friedman et~al.}(2010)\citenamefont{Friedman, Tedesco,
  Campbell, Culbertson, Aifer, Perkins, Myers-Ward, Hite, Eddy, Jernigan
  et~al.}}]{Graphene}
\bibinfo{author}{\bibfnamefont{A.~L.} \bibnamefont{Friedman}},
  \bibinfo{author}{\bibfnamefont{J.~L.} \bibnamefont{Tedesco}},
  \bibinfo{author}{\bibfnamefont{P.~M.} \bibnamefont{Campbell}},
  \bibinfo{author}{\bibfnamefont{J.~C.} \bibnamefont{Culbertson}},
  \bibinfo{author}{\bibfnamefont{E.}~\bibnamefont{Aifer}},
  \bibinfo{author}{\bibfnamefont{F.~K.} \bibnamefont{Perkins}},
  \bibinfo{author}{\bibfnamefont{R.~L.} \bibnamefont{Myers-Ward}},
  \bibinfo{author}{\bibfnamefont{J.~K.} \bibnamefont{Hite}},
  \bibinfo{author}{\bibfnamefont{C.~R.} \bibnamefont{Eddy}},
  \bibinfo{author}{\bibfnamefont{G.~G.} \bibnamefont{Jernigan}},
  \bibnamefont{et~al.}, \bibinfo{journal}{Nano Lett.}
  \textbf{\bibinfo{volume}{10}}, \bibinfo{pages}{3962} (\bibinfo{year}{2010}).

\bibitem[{\citenamefont{He et~al.}(2012)\citenamefont{He, Li, Liu, Guo, Wang,
  Xie, and Wang}}]{TI}
\bibinfo{author}{\bibfnamefont{H.}~\bibnamefont{He}},
  \bibinfo{author}{\bibfnamefont{B.}~\bibnamefont{Li}},
  \bibinfo{author}{\bibfnamefont{H.}~\bibnamefont{Liu}},
  \bibinfo{author}{\bibfnamefont{X.}~\bibnamefont{Guo}},
  \bibinfo{author}{\bibfnamefont{Z.}~\bibnamefont{Wang}},
  \bibinfo{author}{\bibfnamefont{M.}~\bibnamefont{Xie}}, \bibnamefont{and}
  \bibinfo{author}{\bibfnamefont{J.}~\bibnamefont{Wang}},
  \bibinfo{journal}{Appl. Phys. Lett.} \textbf{\bibinfo{volume}{100}},
  \bibinfo{eid}{032105} (pages~\bibinfo{numpages}{3}) (\bibinfo{year}{2012}).

\bibitem[{\citenamefont{Parish and Littlewood}(2005)}]{MeeraPRB}
\bibinfo{author}{\bibfnamefont{M.~M.} \bibnamefont{Parish}} \bibnamefont{and}
  \bibinfo{author}{\bibfnamefont{P.~B.} \bibnamefont{Littlewood}},
  \bibinfo{journal}{Phys. Rev. B} \textbf{\bibinfo{volume}{72}},
  \bibinfo{pages}{094417} (\bibinfo{year}{2005}).

\bibitem[{\citenamefont{Baenninger et~al.}(2005)\citenamefont{Baenninger,
  Ghosh, Pepper, Beere, Farrer, Atkinson, and Ritchie}}]{Matthias}
\bibinfo{author}{\bibfnamefont{M.}~\bibnamefont{Baenninger}},
  \bibinfo{author}{\bibfnamefont{A.}~\bibnamefont{Ghosh}},
  \bibinfo{author}{\bibfnamefont{M.}~\bibnamefont{Pepper}},
  \bibinfo{author}{\bibfnamefont{H.~E.} \bibnamefont{Beere}},
  \bibinfo{author}{\bibfnamefont{I.}~\bibnamefont{Farrer}},
  \bibinfo{author}{\bibfnamefont{P.}~\bibnamefont{Atkinson}}, \bibnamefont{and}
  \bibinfo{author}{\bibfnamefont{D.~A.} \bibnamefont{Ritchie}},
  \bibinfo{journal}{Phys. Rev. B} \textbf{\bibinfo{volume}{72}},
  \bibinfo{pages}{241311} (\bibinfo{year}{2005}).

\bibitem[{\citenamefont{Staring et~al.}(1992)\citenamefont{Staring, van Houten,
  Beenakker, and Foxon}}]{Houten92}
\bibinfo{author}{\bibfnamefont{A.~A.~M.} \bibnamefont{Staring}},
  \bibinfo{author}{\bibfnamefont{H.}~\bibnamefont{van Houten}},
  \bibinfo{author}{\bibfnamefont{C.~W.~J.} \bibnamefont{Beenakker}},
  \bibnamefont{and} \bibinfo{author}{\bibfnamefont{C.~T.} \bibnamefont{Foxon}},
  \bibinfo{journal}{Phys. Rev. B} \textbf{\bibinfo{volume}{45}},
  \bibinfo{pages}{9222} (\bibinfo{year}{1992}).

\bibitem[{\citenamefont{Tripathi and Kennett}(2006)}]{Tripathi06}
\bibinfo{author}{\bibfnamefont{V.}~\bibnamefont{Tripathi}} \bibnamefont{and}
  \bibinfo{author}{\bibfnamefont{M.~P.} \bibnamefont{Kennett}},
  \bibinfo{journal}{Phys. Rev. B} \textbf{\bibinfo{volume}{74}},
  \bibinfo{pages}{195334} (\bibinfo{year}{2006}).

\bibitem[{\citenamefont{Tripathi and Kennett}(2007)}]{Tripathi07}
\bibinfo{author}{\bibfnamefont{V.}~\bibnamefont{Tripathi}} \bibnamefont{and}
  \bibinfo{author}{\bibfnamefont{M.~P.} \bibnamefont{Kennett}},
  \bibinfo{journal}{Phys. Rev. B} \textbf{\bibinfo{volume}{76}},
  \bibinfo{pages}{115321} (\bibinfo{year}{2007}).

\bibitem[{\citenamefont{Neilson and Hamilton}(2010)}]{Neilson}
\bibinfo{author}{\bibfnamefont{D.}~\bibnamefont{Neilson}} \bibnamefont{and}
  \bibinfo{author}{\bibfnamefont{A.~R.} \bibnamefont{Hamilton}},
  \bibinfo{journal}{Phys. Rev. B} \textbf{\bibinfo{volume}{82}},
  \bibinfo{pages}{035310} (\bibinfo{year}{2010}).

\bibitem[{\citenamefont{Ghosh et~al.}(2004)\citenamefont{Ghosh, Pepper, Beere,
  and Ritchie}}]{JPhys_Arindam}
\bibinfo{author}{\bibfnamefont{A.}~\bibnamefont{Ghosh}},
  \bibinfo{author}{\bibfnamefont{M.}~\bibnamefont{Pepper}},
  \bibinfo{author}{\bibfnamefont{H.~E.} \bibnamefont{Beere}}, \bibnamefont{and}
  \bibinfo{author}{\bibfnamefont{D.~A.} \bibnamefont{Ritchie}},
  \bibinfo{journal}{J. Phys.: Condens. Matter} \textbf{\bibinfo{volume}{16}},
  \bibinfo{pages}{3623} (\bibinfo{year}{2004}).

\bibitem[{\citenamefont{Schoonus et~al.}(2008)\citenamefont{Schoonus, Bloom,
  Wagemans, Swagten, and Koopmans}}]{Schoonus}
\bibinfo{author}{\bibfnamefont{J.~J. H.~M.} \bibnamefont{Schoonus}},
  \bibinfo{author}{\bibfnamefont{F.~L.} \bibnamefont{Bloom}},
  \bibinfo{author}{\bibfnamefont{W.}~\bibnamefont{Wagemans}},
  \bibinfo{author}{\bibfnamefont{H.~J.~M.} \bibnamefont{Swagten}},
  \bibnamefont{and} \bibinfo{author}{\bibfnamefont{B.}~\bibnamefont{Koopmans}},
  \bibinfo{journal}{Phys. Rev. Lett.} \textbf{\bibinfo{volume}{100}},
  \bibinfo{pages}{127202} (\bibinfo{year}{2008}).

\bibitem[{\citenamefont{Sun et~al.}(2004)\citenamefont{Sun, Mizuguchi, Manago,
  and Akinaga}}]{Sun}
\bibinfo{author}{\bibfnamefont{Z.~G.} \bibnamefont{Sun}},
  \bibinfo{author}{\bibfnamefont{M.}~\bibnamefont{Mizuguchi}},
  \bibinfo{author}{\bibfnamefont{T.}~\bibnamefont{Manago}}, \bibnamefont{and}
  \bibinfo{author}{\bibfnamefont{H.}~\bibnamefont{Akinaga}},
  \bibinfo{journal}{Appl. Phys. Lett.} \textbf{\bibinfo{volume}{85}},
  \bibinfo{pages}{5643} (\bibinfo{year}{2004}).

\bibitem[{\citenamefont{Ciccarelli et~al.}(2010)\citenamefont{Ciccarelli, Park,
  Ogawa, Ferguson, and Wunderlich}}]{Ciccarelli}
\bibinfo{author}{\bibfnamefont{C.}~\bibnamefont{Ciccarelli}},
  \bibinfo{author}{\bibfnamefont{B.~G.} \bibnamefont{Park}},
  \bibinfo{author}{\bibfnamefont{S.}~\bibnamefont{Ogawa}},
  \bibinfo{author}{\bibfnamefont{A.~J.} \bibnamefont{Ferguson}},
  \bibnamefont{and}
  \bibinfo{author}{\bibfnamefont{J.}~\bibnamefont{Wunderlich}},
  \bibinfo{journal}{Appl. Phys. Lett.} \textbf{\bibinfo{volume}{97}},
  \bibinfo{eid}{082106} (pages~\bibinfo{numpages}{3}) (\bibinfo{year}{2010}).

\bibitem[{\citenamefont{Herring}(1960)}]{Herring}
\bibinfo{author}{\bibfnamefont{C.}~\bibnamefont{Herring}}, \bibinfo{journal}{J.
  Appl. Phys.} \textbf{\bibinfo{volume}{31}}, \bibinfo{pages}{1939}
  (\bibinfo{year}{1960}).

\bibitem[{\citenamefont{Bergman and Stroud}(2000)}]{Bergman}
\bibinfo{author}{\bibfnamefont{D.~J.} \bibnamefont{Bergman}} \bibnamefont{and}
  \bibinfo{author}{\bibfnamefont{D.~G.} \bibnamefont{Stroud}},
  \bibinfo{journal}{Phys. Rev. B} \textbf{\bibinfo{volume}{62}},
  \bibinfo{pages}{6603} (\bibinfo{year}{2000}).

\bibitem[{\citenamefont{Stroud and Pan}(1976)}]{Stroud}
\bibinfo{author}{\bibfnamefont{D.}~\bibnamefont{Stroud}} \bibnamefont{and}
  \bibinfo{author}{\bibfnamefont{F.~P.} \bibnamefont{Pan}},
  \bibinfo{journal}{Phys. Rev. B} \textbf{\bibinfo{volume}{13}},
  \bibinfo{pages}{1434} (\bibinfo{year}{1976}).

\bibitem[{\citenamefont{Guttal and Stroud}(2005)}]{Guttal}
\bibinfo{author}{\bibfnamefont{V.}~\bibnamefont{Guttal}} \bibnamefont{and}
  \bibinfo{author}{\bibfnamefont{D.}~\bibnamefont{Stroud}},
  \bibinfo{journal}{Phys. Rev. B} \textbf{\bibinfo{volume}{71}},
  \bibinfo{pages}{201304} (\bibinfo{year}{2005}).

\bibitem[{\citenamefont{Middleton and Wingreen}(1993)}]{Wingreen}
\bibinfo{author}{\bibfnamefont{A.~A.} \bibnamefont{Middleton}}
  \bibnamefont{and} \bibinfo{author}{\bibfnamefont{N.~S.}
  \bibnamefont{Wingreen}}, \bibinfo{journal}{Phys. Rev. Lett.}
  \textbf{\bibinfo{volume}{71}}, \bibinfo{pages}{3198} (\bibinfo{year}{1993}).

\bibitem[{\citenamefont{Bud'ko et~al.}(1998)\citenamefont{Bud'ko, Canfield,
  Mielke, and Lacerda}}]{Budko}
\bibinfo{author}{\bibfnamefont{S.~L.} \bibnamefont{Bud'ko}},
  \bibinfo{author}{\bibfnamefont{P.~C.} \bibnamefont{Canfield}},
  \bibinfo{author}{\bibfnamefont{C.~H.} \bibnamefont{Mielke}},
  \bibnamefont{and} \bibinfo{author}{\bibfnamefont{A.~H.}
  \bibnamefont{Lacerda}}, \bibinfo{journal}{Phys. Rev. B}
  \textbf{\bibinfo{volume}{57}}, \bibinfo{pages}{13624} (\bibinfo{year}{1998}).

\bibitem[{\citenamefont{Kapitza}(1928)}]{KapitzaBi}
\bibinfo{author}{\bibfnamefont{P.}~\bibnamefont{Kapitza}},
  \bibinfo{journal}{Proc. Roy. Soc. London Ser. A}
  \textbf{\bibinfo{volume}{119}}, \bibinfo{pages}{358} (\bibinfo{year}{1928}).

\bibitem[{\citenamefont{Yang et~al.}(1999)\citenamefont{Yang, Liu, Hong, Reich,
  Searson, and Chien}}]{Yang}
\bibinfo{author}{\bibfnamefont{F.~Y.} \bibnamefont{Yang}},
  \bibinfo{author}{\bibfnamefont{K.}~\bibnamefont{Liu}},
  \bibinfo{author}{\bibfnamefont{K.}~\bibnamefont{Hong}},
  \bibinfo{author}{\bibfnamefont{D.~H.} \bibnamefont{Reich}},
  \bibinfo{author}{\bibfnamefont{P.~C.} \bibnamefont{Searson}},
  \bibnamefont{and} \bibinfo{author}{\bibfnamefont{C.~L.} \bibnamefont{Chien}},
  \bibinfo{journal}{Science} \textbf{\bibinfo{volume}{284}},
  \bibinfo{pages}{1335} (\bibinfo{year}{1999}).

\bibitem[{\citenamefont{Kapitza}(1929)}]{Kapitza}
\bibinfo{author}{\bibfnamefont{P.}~\bibnamefont{Kapitza}},
  \bibinfo{journal}{Proc. Roy. Soc. London Ser. A}
  \textbf{\bibinfo{volume}{123}}, \bibinfo{pages}{292} (\bibinfo{year}{1929}).

\bibitem[{\citenamefont{Johnson et~al.}(2010)\citenamefont{Johnson, Bennett,
  Barua, Lewis, and Heiman}}]{Johnson}
\bibinfo{author}{\bibfnamefont{H.~G.} \bibnamefont{Johnson}},
  \bibinfo{author}{\bibfnamefont{S.~P.} \bibnamefont{Bennett}},
  \bibinfo{author}{\bibfnamefont{R.}~\bibnamefont{Barua}},
  \bibinfo{author}{\bibfnamefont{L.~H.} \bibnamefont{Lewis}}, \bibnamefont{and}
  \bibinfo{author}{\bibfnamefont{D.}~\bibnamefont{Heiman}},
  \bibinfo{journal}{Phys. Rev. B} \textbf{\bibinfo{volume}{82}},
  \bibinfo{pages}{085202} (\bibinfo{year}{2010}).

\bibitem[{\citenamefont{Meir}(1999)}]{Meir}
\bibinfo{author}{\bibfnamefont{Y.}~\bibnamefont{Meir}}, \bibinfo{journal}{Phys.
  Rev. Lett.} \textbf{\bibinfo{volume}{83}}, \bibinfo{pages}{3506}
  (\bibinfo{year}{1999}).

\end{thebibliography}
\end{document}

\title {Supplementary material: Colossal non-saturating linear magnetoresistance in two-dimensional electron systems at a GaAs/AlGaAs heterointerface}

\author{M. A. Aamir}
\email{aamir@physics.iisc.ernet.in}
\affiliation{Department of Physics, Indian Institute of Science, Bangalore 560 012, India.}
\author{Srijit Goswami}
\affiliation{Department of Physics, Indian Institute of Science, Bangalore 560 012, India.}
\author{Matthias Baenninger}
\affiliation{Cavendish Laboratory, University of Cambridge, J.J. Thomson Avenue, Cambridge CB3 0HE, United Kingdom.}
\author{Vikram Tripathi}
\affiliation{Department of Theoretical Physics, Tata Institute of Fundamental Research, Homi Bhabha Road, Mumbai 400005, India}
\author{Michael~Pepper}
\affiliation{Department of Electrical and Electronic Engineering, University College, London WC1E 7JE, United Kingdom}
\author{Ian~Farrer}
\affiliation{Cavendish Laboratory, University of Cambridge, J.J. Thomson Avenue, Cambridge CB3 0HE, United Kingdom.}
\author{David~A.~Ritchie}
\affiliation{Cavendish Laboratory, University of Cambridge, J.J. Thomson Avenue, Cambridge CB3 0HE, United Kingdom.}
\author{Arindam~Ghosh}
\affiliation{Department of Physics, Indian Institute of Science, Bangalore 560 012, India.}

\maketitle

\section{Analysis of the $I-V_{sd}$ characteristics}

Fig.~1a shows the gate voltage dependence of the $I-V_{sd}$ characteristics in Region II (as defined in the main text) at about 8 K. As is clear, the threshold $V_T$, which emerges in the $I-V_{sd}$ characteristics in the region II, has a distinct gate voltage dependence (Fig.~1b). This is a remarkably linear dependence which does not break down up to the lowest gate voltages that we have applied. This linear dependence was also observed in other devices in a different wafer (not shown). The voltage threshold seems to be associated with the Coulomb blockade of the puddles of electrons, which is expected to form at such negative gate voltages. It is not clear how the linear dependence emerges, but may be the outcome the manner in which gate voltage changes the voltage barrier between these puddles of electrons.

\begin{figure}[h]

\includegraphics[scale=0.23]{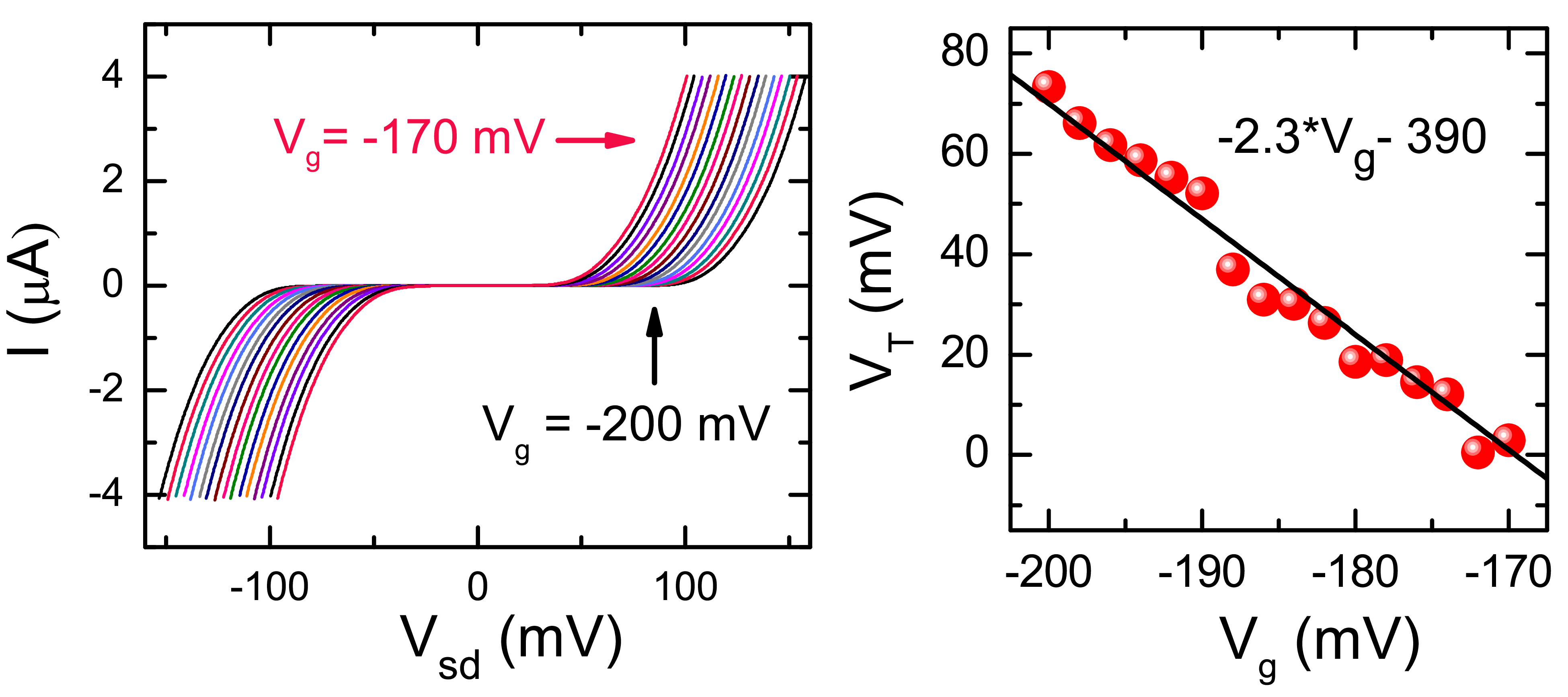}

\caption{(a) Gate voltage dependence of the $I-V_{sd}$ characteristics of the device in Region II from $V_g=-170$ mV to $V_g=-200$ mV in steps of 2 mV (b) Threshold voltage $V_T$ as a function of gate voltage extracted from the $I-V_{sd}$ characteristics in (a)}

\end{figure}

The fact that the system under study is highly inhomogeneous can be further ascertained by analysing the non-linear nature of $I-V_{sd}$ characteristics. In a typical $I-V_{sd}$ curve, the current rises rapidly after having crossed the threshold $V_T$. The $I-V_{sd}$ characteristics, in fact, display a power law with the form $I=A(V_{sd}-V_{T})^\zeta$. Fig. 2a shows this for a few magnetic fields, at 4.9 K. Interestingly, very similar $I-V_{sd}$ curves have been observed in a disordered array of quantum dots~\cite{Wingreen, Durouz, Rimberg, Partha_structural, Reichhardt, Partha, ourpaper}. There, the bias threshold is directly related to the Coulomb gap of the series of quantum dots that define the electron trajectories from one lead to the other. $\zeta$ depends on the type and strength of disorder, as well as the geometry of the device. It is not surprising that such $I-V_{sd}$ curves appear in our devices as well, since the 2DES is also expected to form a disordered array of electron puddles in the pinched-off regime at very negative gate voltages~\cite{Tripathi06,Tripathi07,Neilson}.

\begin{figure}[h]

\includegraphics[scale=0.23]{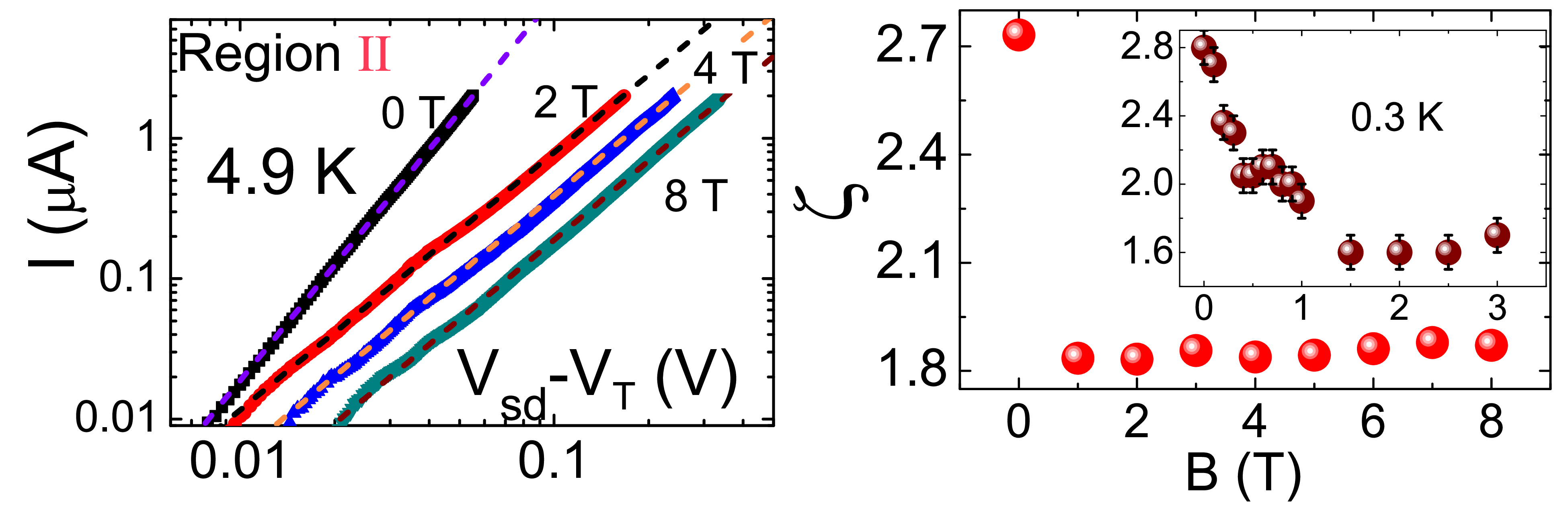}

\caption{(a)$I-V_{sd}$ characteristics (with threshold $V_T$ subtracted) cast in the log-log plot for magnetic fields 0~T, 2~T, 4~T and 8~T at 4.9 K. Dashed lines represent power law fitting of the respective curve (b) Plot of $\zeta$ as a function of magnetic field B for measurements presented in (a). Inset: A finer scan of $\zeta$ vs B (at 0.3~K) for another cool-down}

\end{figure}

Interestingly, the value of $\zeta$ drops sharply on the application of the magnetic field (Fig. 2b). At zero field, $\zeta=2.8$, and falls to $\zeta=1.8$ by 1 T and saturates thereafter. A finer scan in another cool-down (inset of Fig. 2b) confirms that this behavior is indeed reproducible. The high $\zeta=2.8$ that we observe at $B=0$~T is close to values reported for array of quantum dots with structural disorder~\cite{Partha_structural}, whereas $\zeta \simeq 1.6$ at higher $B$ is consistent with Middleton-Wingreen model of a capacitively coupled quantum dot array with on-site charge offset disorder~\cite{Wingreen}. The reason for this transition of $\zeta$ with $B$ is unclear, but may be connected to redistribution of the charge inhomogeneity due to wavefunction squeezing in the puddles. Nonetheless, the $I-V_{sd}$ traces indicate that at sub-pinch-off gate voltages the 2DES is highly inhomogeneous, and may be visualised as an array of quantum dots with large structural and size disorder.

\section{Bias dependence of NLMR}
\begin{figure}[h]

\includegraphics[scale=0.15]{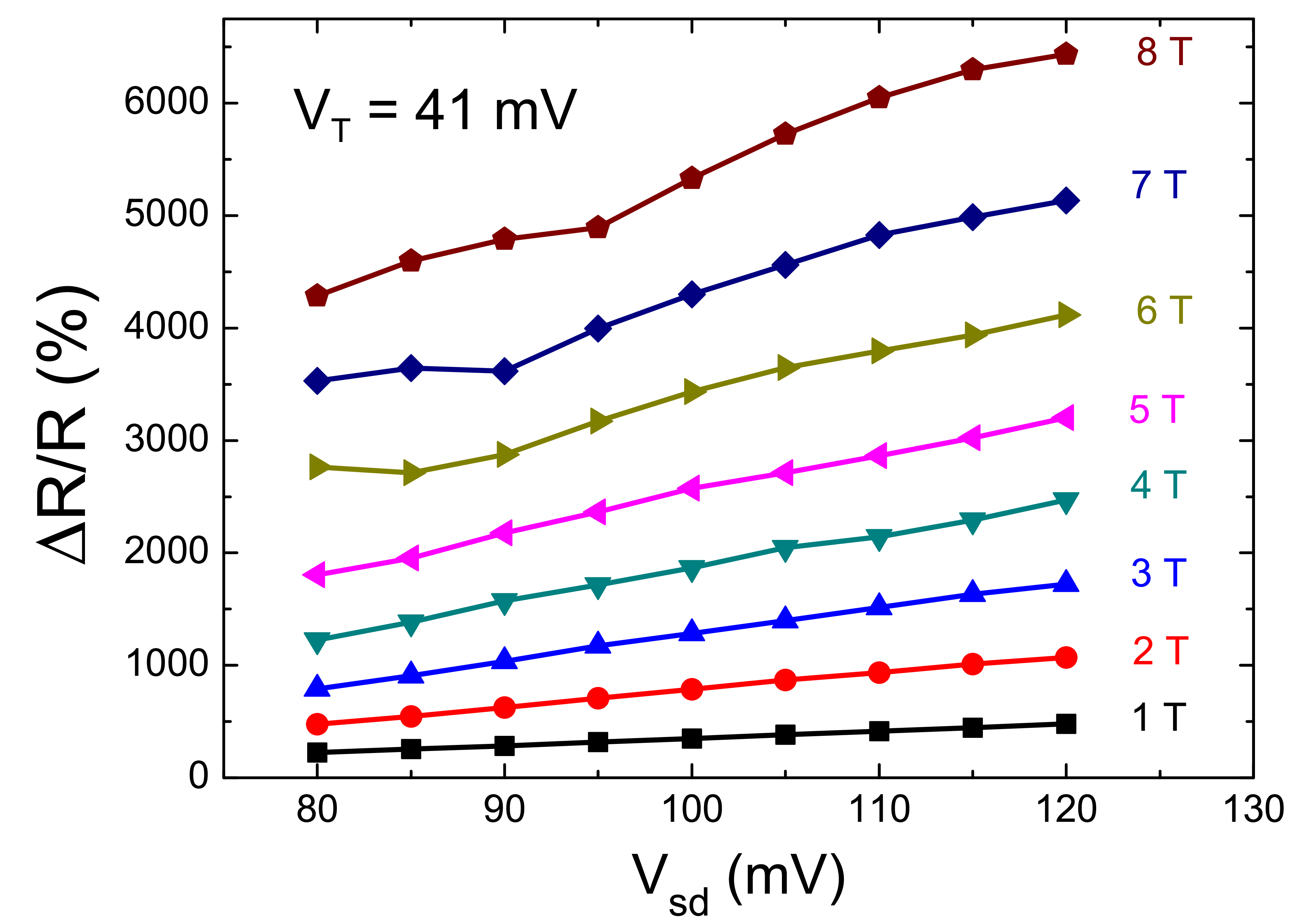}

\caption{$\Delta R/R$ as a function of $V$ for magnetic fields from 1 T to 8 T at 4.9 K}

\end{figure}

The fact that $\zeta$ drops significantly from 2.8 (at $B=0$) to 1.8 (at $B=0$, and thereafter) implies that beyond the voltage threshold, $I$ increases much faster when the magnetic field is zero. In other words, with increasing $V_{sd}$, $R(B=0)$ drops faster than $R(B>0)$. As a result, $\Delta R/R$ (as defined in the main text) is an increasing function of $V_{sd}$. Fig.~3 shows the variation of MR with $V_{sd}$ for all the magnetic fields from 1~T to 8~T.
%The sharp difference in $\zeta$ between $I-V_{sd}$ curve at 0~T and that of all other magnetic fields also give rise to a strong MR dependence on $V_{sd}$. This is because the MR is obtained by normalizing the resistance at other fields with respect to that at 0~T. The resistance at $B=0$~T drops much faster than resistance at other magnetic fields with increasing $V_{sd}$ and hence the MR also rises with increasing $V_{sd}$. Fig. 3 shows the plot of MR against $V_{sd}$ for all the magnetic fields from 1~T to 8~T in steps of 1~T.

%The high $\zeta$=2.8 that we observe at zero magnetic field is very close to values reported for array of quantum dots with structural disorder~\cite{Partha_structural}. The fact that magnetic field rapidly reduces $\zeta$ suggesting reduction in the degree of disorder may have the following two explanations. The disorder may have reduced because magnetic field deviates all the current paths, thus channelizing all the current paths in a more systematic way. Another possibility is that the magnetic field squeezes the states in the dense regions of the medium effectively reducing size and structural disorder. However, a more quantitative analysis of this result is need to shed more light on this process behind the this phenomenon.

\section{Linear magnetoresistance in other devices}

\begin{figure}[h]

\includegraphics[scale=0.27]{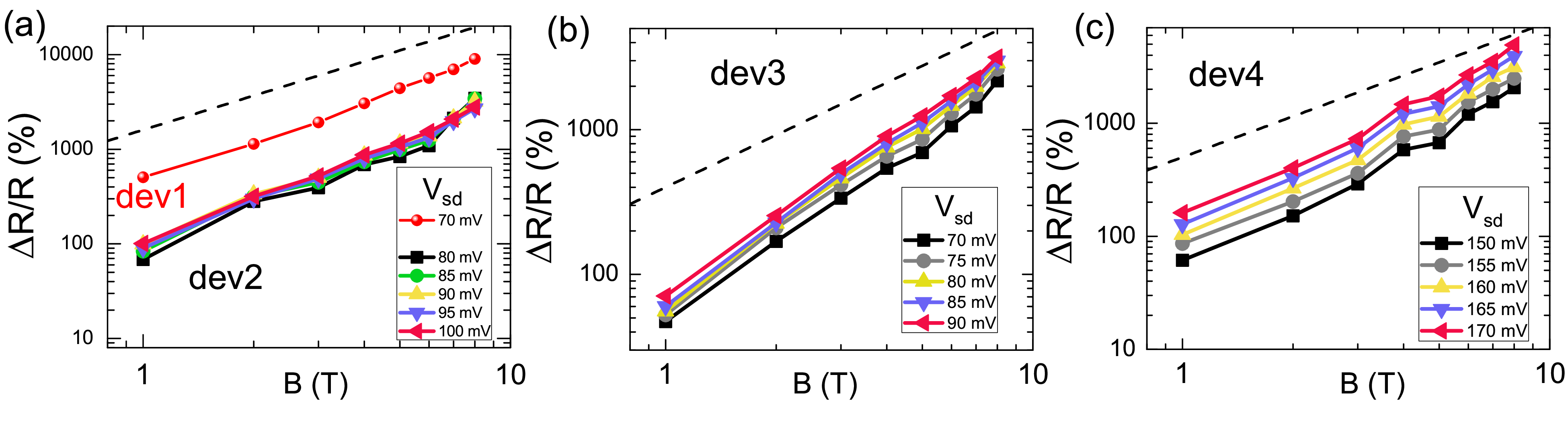}

\caption{(a) Comparison of linear MR of the dev1 (reported in the paper) with that of dev2 (b) dev3 (c) dev4. In all plots, the dashed line is a power law curve with exponent 1.2}

\end{figure}

Fig.~4 presents $\Delta R/R$ (calculated in the same way as in the main text) for other devices dev2, dev3 and dev4 in another GaAs/AlGaAs heterostructure wafer. Fig.~4a compares one of the MR curves for dev1 (from Fig.~3b of the main text) with those of dev2. The form of the MR for dev2 is same as dev1 and follows a power law with exponent 1.2 (dashed line), thus exhibiting a near linear dependence. Fig.~4b and Fig.~4c show MR from two devices which show a slightly different exponent. %A closer analysis shows that these approximately follow $MR \propto B^{1.5}$.

\section{Theoretical foundation of NLMR in an inhomogeneous medium}

\begin{figure}[h]

\includegraphics[scale=0.7]{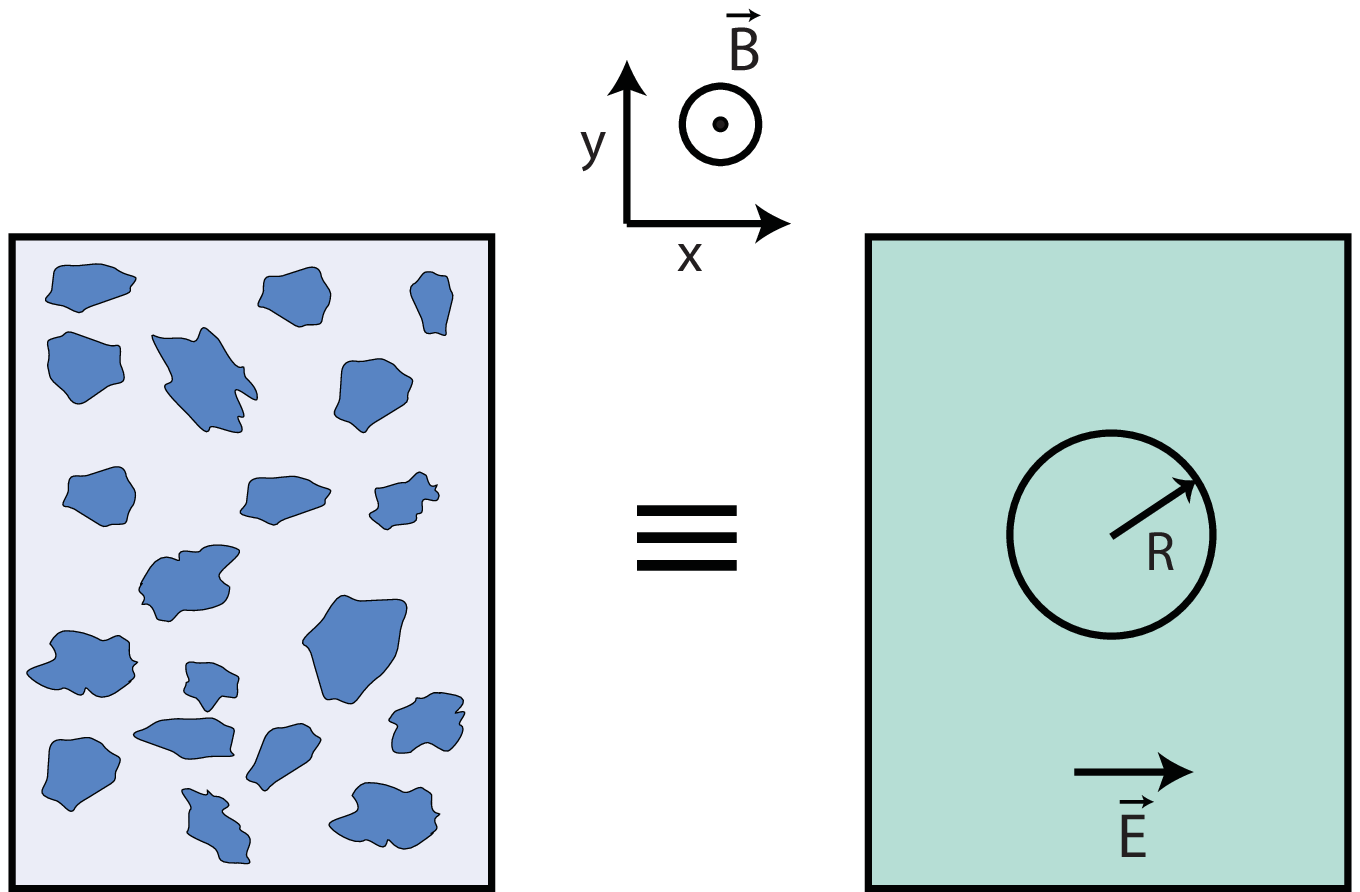}

\caption{Schematic describing the model used to understand the NLMR: Left: A two-dimensional conducting medium is dispersed inhomogeneously with another material. Right: An effective medium equivalent to the one on the left.}

\end{figure}

In order to understand the linear magnetoresistance in an inhomogeneous medium, we adapt the 3D magnetoresistance analysis of Stachowiak~\cite{Stachowiak} (itself motivated by Herring's effective medium treatment~\cite{Herring}) to 2D systems. Consider a two-dimensional conducting material in the $x-y$ plane (Fig.~5) with longitudinal and Hall conductivities $\sigma_{11}$ and $\sigma_{12}$ respectively in presence of a magnetic field $H_{z}$ in the $z-$direction. Suppose we make another 2D conductor by inhomogeneously dispersing this material. The local conductivity of this composite material is $\sigma=\left(\begin{array}{cc} \sigma_{11} & \sigma_{12}\\ -\sigma_{12} & \sigma_{11}\end{array}\right).$ We denote the corresponding effective conductivity of the composite by $\tilde{\sigma}=\left(\begin{array}{cc} \tilde{\sigma}_{11} & \tilde{\sigma}_{12}\\
-\tilde{\sigma}_{21} & \tilde{\sigma}_{11}\end{array}\right).$ We would like to obtain a relation between the local and effective
conductivity. The measured resistivity of the sample is given by $\tilde{\rho}=\tilde{\sigma}_{11}/(\tilde{\sigma}_{11}^{2}+\tilde{\sigma}_{12}^{2}).$
\\
\\
Consider a small circular region of radius $R$ made of the conducting material and embedded in an effective medium formed of the composite. Let there be a uniform electric field $E_{0x}$ along the $x-$direction far from this disc. The resulting current $\mathbf{j}$ everywhere satisfies the continuity equation, $\nabla\cdot\mathbf{j}=0.$ Using Ohm's law, $\mathbf{j}=\tilde{\sigma}\mathbf{E}$ etc., together with the definition $\mathbf{E}=-\nabla\psi,$ where $\psi$ is the electrostatic potential, it is easily seen that the Laplace equation $\nabla^{2}\psi=0$ is satisfied. We need the solution for $\psi$ subject to the following boundary conditions: (a) $\lim_{r\rightarrow\infty}\nabla\psi=-(E_{0x},0),$ (b) $\psi(r)$ is continuous at $r=R$ and (c) the normal component of the current at $r=R,$ $\mathbf{j}\cdot\hat{\mathbf{r}},$ is continuous.

Solving the Laplace equation with these boundary conditions, one obtains the following solution for the potential $\psi_{i}$ inside the disc:

\begin{align}
\psi_{i}(x,y)&=C_{i1}x+D_{i1}y,\nonumber \\
C_{1i}&=-\frac{3\tilde{\sigma}_{11}(2\tilde{\sigma}_{11}+\sigma_{11})E_{0x}}{(2\tilde{\sigma}_{11}+\sigma_{11})^{2}+(\tilde{\sigma}_{12}-\sigma_{12})^{2}},\nonumber \\
D_{1i}&=\frac{3\tilde{\sigma}_{11}(\tilde{\sigma}_{12}-\sigma_{12})E_{0x}}{(2\tilde{\sigma}_{11}+\sigma_{11})^{2}+(\tilde{\sigma}_{12}-\sigma_{12})^{2}}.\label{eq:psi-int}
\end{align}

Now we impose self-consistency, i.e., spatially averaging the internal field obtained above should give us the macroscopic electric fields prevailing in the effective medium. Thus imposing $\langle E_{ix}\rangle=E_{0x}$ and $\langle E_{iy}\rangle=E_{0y}=0$ we have

\begin{align}
\left\langle \frac{3\tilde{\sigma}_{11}(2\tilde{\sigma}_{11}+\sigma_{11})}{(2\tilde{\sigma}_{11}+\sigma_{11})^{2}+(\tilde{\sigma}_{12}-\sigma_{12})^{2}}\right\rangle  & =1,\label{eq:selfcon-x}\\
\left\langle \frac{3\tilde{\sigma}_{11}(\tilde{\sigma}_{12}-\sigma_{12})}{(2\tilde{\sigma}_{11}+\sigma_{11})^{2}+(\tilde{\sigma}_{12}-\sigma_{12})^{2}}\right\rangle  & =0.\label{eq:selfcon-y}\end{align}
For classical ohmic conductors,

\begin{align*}
\sigma_{11} & =\frac{\sigma_{0}}{1+(H_{z}/H_{0})^{2}},\\
\sigma_{12} & =\frac{\sigma_{0}(H_{z}/H_{0})}{1+(H_{z}/H_{0})^{2}},
\end{align*}

where $H_{0}^{-1}=e\tau/m$ is related to the mean free scattering time $\tau$ for the electrons in the material, and $\sigma_{0}=ne^{2}\tau/m$ is the usual Drude conductivity. When $H_{z}\gg H_{0},$ $\sigma_{12}\sim\sigma_{0}(H_{0}/H_{z})\gg\sigma_{11}\sim\sigma_{0}(H_{0}/H_{z})^{2}.$ This gives us a clue for the origin of linear magnetoresistance: in the inhomogeneous medium, the effective longitudinal conductivity takes on a bit of the local transverse conductivity.

Consider now Eq. \ref{eq:selfcon-y}. The solution is $\langle\sigma_{12}\rangle=\tilde{\sigma}_{12}.$ Next in Eq. \ref{eq:selfcon-x}, consider the strong magnetic field case first where we expect $\tilde{\sigma}_{11}\gg\sigma_{11}$ on account of the admixture of $\sigma_{12}$ in the effective longitudinal
conductivity. This simplifies Eq. \ref{eq:selfcon-x} to

\begin{align*}
\left\langle1-\left(\frac{\tilde{\sigma}_{12}-\sigma_{12}}{2\tilde{\sigma}_{11}}\right)^{2}\right\rangle & \approx\frac{2}{3},
\end{align*}

which gives us

\begin{align}
\tilde{\sigma}_{11} & \approx\sqrt{\frac{4}{3}\langle(\tilde{\sigma}_{12}-\sigma_{12})^{2}\rangle}\sim O(1/H_{z}).\label{eq:lin-magnetores}
\end{align}

Correspondingly, the effective longitudinal resistivity $\tilde{\rho}=\tilde{\sigma}_{11}/(\tilde{\sigma}_{11}^{2}+\tilde{\sigma}_{12}^{2})\sim O(H_{z}).$
For weak magnetic fields, the effective longitudinal conductivity will be close to $\sigma_{11}$ and the magnetoresistance will go as $H_{z}^{2}.$